\documentclass[useAMS,usenatbib]{mn2e}
\usepackage{amssymb}
\usepackage{graphicx}
\begin{document}
\title[Diffractive Microlensing I]{Diffractive Microlensing I:
  Flickering Planetesimals at the Edge of the Solar System}
\date{Accepted 2009 November 26.  Received 2009 November 23; in
  original form 2009 October 20}
\author[J. Heyl]{Jeremy Heyl\\
Department of Physics and Astronomy, University of British Columbia,
Vancouver, British Columbia, Canada, V6T 1Z1;\\
 Email: heyl@phas.ubc.ca; Canada Research Chair}
\pagerange{\pageref{firstpage}--\pageref{lastpage}} \pubyear{2009}

\maketitle

\label{firstpage}

\begin{abstract}
  Microlensing and occultation are generally studied in the geometric
  optics limit.  However, diffraction may be important when recently
  discovered Kuiper-Belt objects (KBOs) occult distant stars.  In
  particular the effects of diffraction become more important as the
  wavelength of the observation and the distance to the KBO increase.
  For sufficiently distant and massive KBOs or Oort cloud objects not
  only is diffraction important but so is gravitational lensing.  For
  an object similar to Eris but located in the Oort cloud, the
  signature of gravitational lensing would be detected easily during
  an occultation and would give constraints on the mass and radius of
  the object.
\end{abstract}

\section{Introduction}

\citet{Bailey:1976p1677} first argued that small bodies in the distant
solar system could be detected through stellar occultations.
\citet{1987AJ.....93.1549R} developed the treatment of occultation by
irregular bodies including diffraction.  More recently with the
discovery of the population of Kuiper Belt objects
\citep{Kuiper:1951p1773,Jewitt:1999p1844}, several research groups
have begun searching for more distant and smaller bodies through
occultations
\citep{Roques:2000p1661,2008AJ....135.1039B,Roques:2009p1666}.
Because Kuiper Belt objects (KBOs) typically subtend small angles, the
diffraction of radiation around the objects may be important even at
visible wavelengths and naturally more important at longer wavelengths
\citep{Roques:6p1676}.  The discovery of more distant and more massive
KBOs such as Eris \citep{2005ApJ...635L..97B} begs the question of
whether gravitational lensing of background stars by large KBOs and
objects in the Oort cloud \citep{Oort:1950p1427} is important.
\citet{2002astro.ph..9545C} argued that for distant massive KBOs and
Oort cloud objects lensing may be important; furthermore,
\citet{2005ApJ...635..711G} argued that GAIA could measure the
astrometric displacement from microlensing by an planet more massive
than a few Jupiters within $10^4$~AU regardless of its location on the
sky.  The technique in this paper probes much lower mass objects, but
also exploits lensing to provide constraints on the properties of the
asteroid.
 
Generally the diffractive effects of microlensing are neglected
because the variation in the time delays across the lens is usually
much larger than the coherence time of the observation, $1/\Delta \nu$
where $\Delta \nu$ is the bandwidth of the observation.  However, near
a caustic crossing, diffraction may be important as argued by
\citet{1995ApJ...455..443J} to account for rapid variations in the
light from Q2237+0305 (The Einstein Cross).  Typically the
differential time delay highly magnified images for a point lens is
about $2 GM/c^3$, the crossing time over the Schwarzschild radius of
the lens; consequently, for the diffractive effects of lensing to be
observable the Schwarzschild radius of the lens should be comparable
or larger than the wavelength of the radiation.  The largest of the
Kuiper Belt objects, Eris, has a Schwarzschild radius $R_S=2GM/c^2
\approx 25 \mu$m.  Therefore, quite naturally for observations of
large KBOs diffractive microlensing may be important for observations
in the near and mid-infrared whenever the gravitational effects are
important.  This paper focuses on just this regime.

The first section, \S\ref{sec:diffr-micr}, outlines microlensing in
the diffractive regime and generalizes the earlier results to include
occultation.  This yields a expression for the transmission that is
nearly identical to the unlensed result.  The next section,
\S\ref{sec:results}, outlines the types of objects for which
diffractive lensing may be important, examines several interesting
cases and connects the diffraction patterns to the geometric limit
(\S\ref{sec:geometric-optics}) in the limit where the lensing is weak.
The final section (\S\ref{sec:conclusions}) outlines how diffractive
lensing could constrain the properties of known objects and speculates
on the probability of such lensing events.

\section{Diffractive Microlensing}
\label{sec:diffr-micr}

The following definitions will prove useful throughout the paper.
The source lies a distance $v$ in the plane of the sky from the centre
of the lensing occulter.  The variable $u$ is a radial variable over
the plane containing the lensing occulter and $\varphi$ gives the
polar angle in this plane with $\varphi=0$ pointing toward the
projection of the source onto this plane The magnification is given by
squared modulus of the integral over the phases in the lens plane
\citep{1992grle.book.....S}
\begin{equation}
\mu_\omega =
\left | \int_{u_d}^\infty du  u^{1-if} e^{iu^2/2} 
\int_0^{2\pi} d\varphi  e^{-ifuv\cos \varphi} \right |^2.
\label{eq:1}
\end{equation}
The lower bound accounts for the effects of occultation with
\begin{equation}
u_d = r_d \sqrt{\frac{\omega_d}{c} \frac{D_s}{D_d
    D_{ds}}}. \label{eq:6}.
\end{equation}
where $r_d$  is the radius of the occulter and $\omega_d$ is the angular frequency of the radiation at
the lensing occulter. 
The value $u_d$ is related to the
Fresnel number such that $F=u_d^2/(2\pi)$.  The
other parameter is 
$f = 2 R_S \omega_d/c$ ($R_S=2 G M_d/c^2$) where
$M_d$ is the mass of the occulter.
The Einstein radius is the characteristic
length of the lens,
\begin{equation}
R_E = \sqrt{2 R_S \frac{D_d D_{ds}}{D_s} }=\frac{\sqrt{f}}{u_d} r_d. \label{eq:3}
\end{equation}
Performing the angular integral yields
\begin{equation}
\mu_\omega =  
\left | \int_{u_d}^\infty u^{1-if} e^{iu^2/2} J_0(uv) du \right |^2.\label{eq:9}
\end{equation}
The limit where the gravitational field of the lens is negligible is
$f=0$, so the effect of gravity on the form of the integral is quite
modest.

\subsection{Evaluating the Integral}
\label{sec:evaluating-integral}

The integral can be calculated in closed form in terms of the
confluent hypergeometric function ($_1F_1(a;b;z)$) for $u_d=0$.
Using relation (6.631.1) from \citet{Grad94} yields
\begin{eqnarray}
\int_0^\infty u^{1-if} e^{iu^2/2} J_0(uv) du \!\!&=&\!\!e^{\pi f/4} 
e^{i \left ( \pi - f \ln 2\right)/2}
\Gamma \left (1 - i \frac{f}{2}\right ) \times \nonumber \\
& & ~{}_1F_1 \left ( 1 - i \frac{f}{2}; 1 ; -i\frac{v^2}{2} \right )\label{eq:11}.
\end{eqnarray}
The result for $f=0$ is simply $i\exp(-iv^2/2)$.
These analytic results are used to calculate the integral in
Eq.~(\ref{eq:9}), both with and without lensing.  Rather than
performing the integration to $u\rightarrow \infty$, the result of the
integration from $u=0$ to $u=u_d$ is subtracted from Eq.~\ref{eq:11}.

\subsection{Geometric Optics}
\label{sec:geometric-optics}

To compare with the standard, high frequency results for lensing and
occultation \citep[e.g.][]{Agol:2002p1710}, the geometric limit is
useful.  In the geometric limit the images are located at those
places in the image plane $u$ where the phase in the integral for
$\mu_\omega$ is stationary,
so $v=u-f/u$, yielding two image positions,
\begin{equation}
u_\pm = \frac{1}{2} \left ( v \pm \sqrt{v^2 + 4 f} \right ), \phi=0\label{eq:14}
\end{equation}
with the magnifications
\begin{equation}
\mu_\pm = \frac{u_\pm}{v} \left |\frac{d u_\pm}{d v} \right |= \frac{1}{2} \left ( \frac{v^2+2 f}{v\sqrt{v^2+4 f}} \pm 1 \right ).\label{eq:15}
\end{equation}
The size of the Einstein radius in these units is simply $\sqrt{f}$,
so if $\sqrt{f} \gtrsim u_d$ one expects that lensing will be
important.  To account for the occultation in the limit of a point
source, the only the magnification for images with $u_\pm>u_d$ is
included \citep{Agol:2002p1710}. As the following section will show,
the geometric optics approximation not only provides an estimate of
the magnification but it also provides a mapping to convert an
unlensed diffraction pattern to an approximation of the lensed pattern
through the locations of images in Eq.~(\ref{eq:14}).


\section{Results}
\label{sec:results}

Because Kuiper Belt objects provide the inspiration for this work, it
is natural to take the largest KBO, Eris, as an exemplar.  Eris is a
dwarf planet orbiting the Sun at a distance of up to 100 AU.  Its
radius is about 1200 km
\citep{2006ApJ...643L..61B,2008ssbn.book..161S}, yielding a value of
$u_d=770$ at one micron and a Frensel number of about $10^5$.  Its
mass is about $1.7 \times 10^{25}$~g \citep{2007Sci...316.1585B} so
its Schwarzschild radius is about 25 microns so $f\approx 310$ at one
micron.  At one hundred AU, the diffractive effects are not important
over a reasonable bandwidth.  Furthermore, $u_d \gg \sqrt{f}$ so
microlensing is not important either.

\begin{figure}
\includegraphics[width=3.4in]{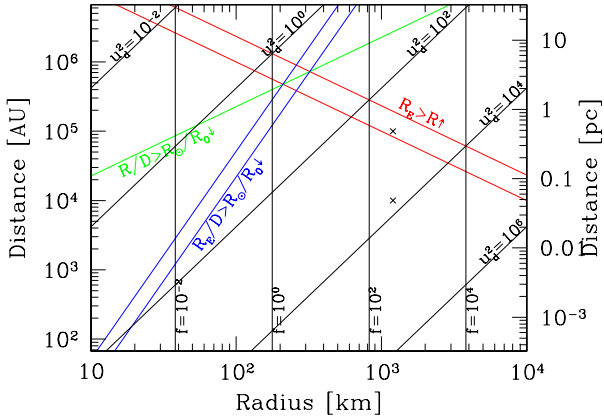}
\caption{The values of $f$ and $u_d^2$ as a function of the radius of
  the asteroid and its distance for $\lambda=1\mu{\rm m}$.  Both $f$
  and $u_d^2$ scale as $\lambda^{-1}$.  The green line $R/D>R_\odot/R_0$
  indicates the region where the angular size of the asteroid exceeds
  that of a solar-type star in the bulge, so the point-source
  approximation for occultation is valid. The blue line
  $R_E/D>R_\odot/R_0$ indicates the region where the angular size of
  the Einstein radius of the asteroid exceeds that of a solar-type
  star in the bulge, so the point-source approximation for
  microlensing is valid.  The red line $R_E>R$ indicates the region where
  the Einstein radius is larger than the radius of the object so
  lensing will dominate over occultation.  In the latter two cases
  the density of the asteroid is assumed to equal that of Eris.  A
  second line to the left gives the boundary for an asteroid 2.3
  times denser similar to the density of Earth. The upper cross
  indicates the parameters for Figs.~\ref{fig:1E5AU_Zero}
  and~\ref{fig:1E5AU_0.2}.  The lower cross gives the parameters for
  Fig.~\ref{fig:1E4AU}.}
\label{fig:regimes}
\end{figure}

On the other hand, Eris is likely to be one of the closest and biggest
members of a large population of KBOs and Oort cloud objects.  Eris
provides a fiducial density of about 2.3 g cm$^{-3}$ (about one-half
that of Earth, 5.5 g cm$^{-3}$) to examine a variety of objects at
various distances and sizes and determine the possible regimes where
microlensing and diffraction could both be important.
Fig.~\ref{fig:regimes} examines the various regimes using the fiducial
density of Eris and a wavelength of one micron.  The values of $f$ and
$u_d^2$ at one micron are depicted on the figure, and they both scale
inversely with wavelength.  In particular to exploit the point-source
approximation in the previous section, the angular size of the object
or the angular size of the object's Einstein radius should be greater
than the angular size of the typical source (for example, a solar-type
star in the bulge taken to be 7.6~kpc away).  In the figure the
point-source approximation is valid for objects that lie either to the
right of the green line or the blue lines (the left-hand blue line
uses Earth's mass density).  Microlensing becomes important for
objects near or above the red lines (the lower red line uses Earth's
mass density).

The figure indicates that observations of Eris at the distance of
100~AU give $u_d^2\approx 10^6$ at one micron and $f\approx 10^{2.5}$,
so $u_d^2\gg f \gg 1$ so neither lensing nor diffraction are important
for Eris, as discussed earlier (at least at one micron --- Eris will
diffract for $\lambda \gtrsim 1$m).  On the other hand, if there were
an Eris-like object at $10^5$~AU (upper cross) and it were observed at
ten microns, $u_d=7.8$ and $f\approx 31.4$, yielding interesting
diffractive effects over a reasonable bandwidth.  At this distance
Eris would subtend about $1.7 \times 10^{-10}$ radians, larger than
any sun-like star beyond a few hundred parsecs, so the star can be
taken to be a point source.  At $10^5$~AU the asteroid is essentially
beyond the realm of the solar system as passing molecular clouds would
unbind objects beyond $4 \times 10^{4}$~AU \citep{1983MNRAS.204..603B}.
Because this is a statistical process, some objects would remain but
they would be quite rare.

Regardless of the paucity of objects at this distance, an Eris-like
object at $10^5$~AU provides an excellent illustation of diffractive
microlensing and occultation.  Beyond a distance of one parsec, Eris
would no longer fully occult a star because even when fully aligned
the lensing would be sufficient to bend the light around the limb of
Eris and toward Earth.  Even at this distance solar-like stars beyond
a kiloparsec would be essentially point sources for the purposes of
the lensing signature.  Furthermore, the diffraction pattern depends
on the values of $u_d$ and $f$, so it would also apply to more
massive and nearer objects but at larger wavelengths --- for example,
observations of an object like Earth at 85~GHz from a distance of
8,000~AU give the same set of values.

\begin{figure}
\includegraphics[width=3.4in]{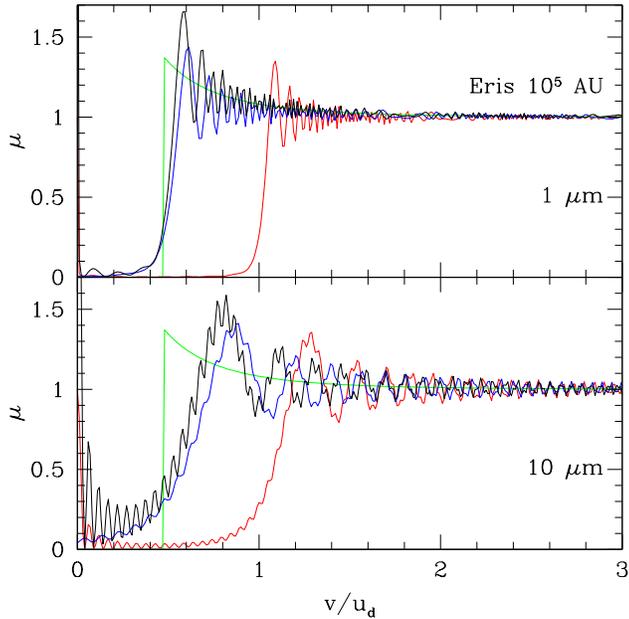}
\caption{The magnification $\mu$ as a function of the location of the
  source ($v$) for an Eris-like object at $10^5$~AU about 0.5~pc.  The
  value of $v=u_d$ just as the source passes behind the edge of the
  occulter.  For $v=0$ the source lies directly behind the centre of
  the circular occulter.  The black curve gives the magnification
  including gravitational lensing.  The red curve gives the standard
  Fresnel diffraction result.  The green curve gives the microlensing
  magnification in the geometric optics limit, and the blue curve
  gives the lensed version of the unlensed diffraction pattern (see
  text).  The observations lie at one and ten microns and have zero
  bandwidth.  The upper panel gives $\lambda=1\mu$m ($f=314$ and
  $u_d=25.$) and the lower panel gives $\lambda=10\mu$m ($f=31.4$
  and $u_d=7.8$).}
\label{fig:1E5AU_Zero}
\end{figure}

Although such objects would be difficult to detect and the possibility
of an occultation event for a given object would be small, the
properties of an occultation and microlensing event for an Eris-like
object at $10^4$ and $10^5$~AU are extremely illustrative.  At
$10^5$~AU, diffraction is important even at one micron.
Fig.~\ref{fig:1E5AU_Zero} gives the diffraction and lensing pattern
for an Eris-like asteroid at $10^5$~AU at one and ten microns.  The
complete diffraction pattern including the effects of gravitational
lensing is given in black.  The diffraction pattern with $f=0$ (no
lensing) is given in red.  The diffraction pattern including lensing
oscillates about the geometric-optics value of the microlensing
magnification (green curve) from Eq.~(\ref{eq:15}).  One can obtain an
approximate idea of the diffraction pattern including lensing by using
the lensing mapping {\em i.e.} $v \rightarrow v - f/v$ and taking the
product of the geometric magnification with the unlensed diffraction
pattern resulting in the blue curve.  When the lensing effect is weak
as in Fig.~\ref{fig:1E4AU}, this {\em a posteriori} lensing correction
works quite well, so occultation diffraction patterns can be corrected
for a modest amount of lensing without calculating the full lensing
diffraction pattern.  However, this mapping fails to reproduce the
central fringing that occurs when both diffraction and lensing are
important.

\begin{figure}
\includegraphics[width=3.4in]{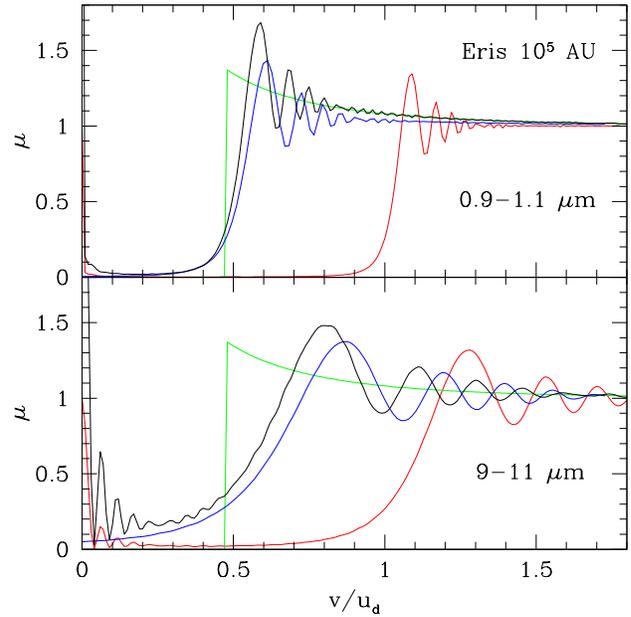}
\caption{Same as Fig~\ref{fig:1E5AU_Zero} but for a twenty-percent bandwidth.}
\label{fig:1E5AU_0.2}
\end{figure}

\begin{figure}
\includegraphics[width=3.4in,clip,trim = 0 0.5in 0 0.2in]{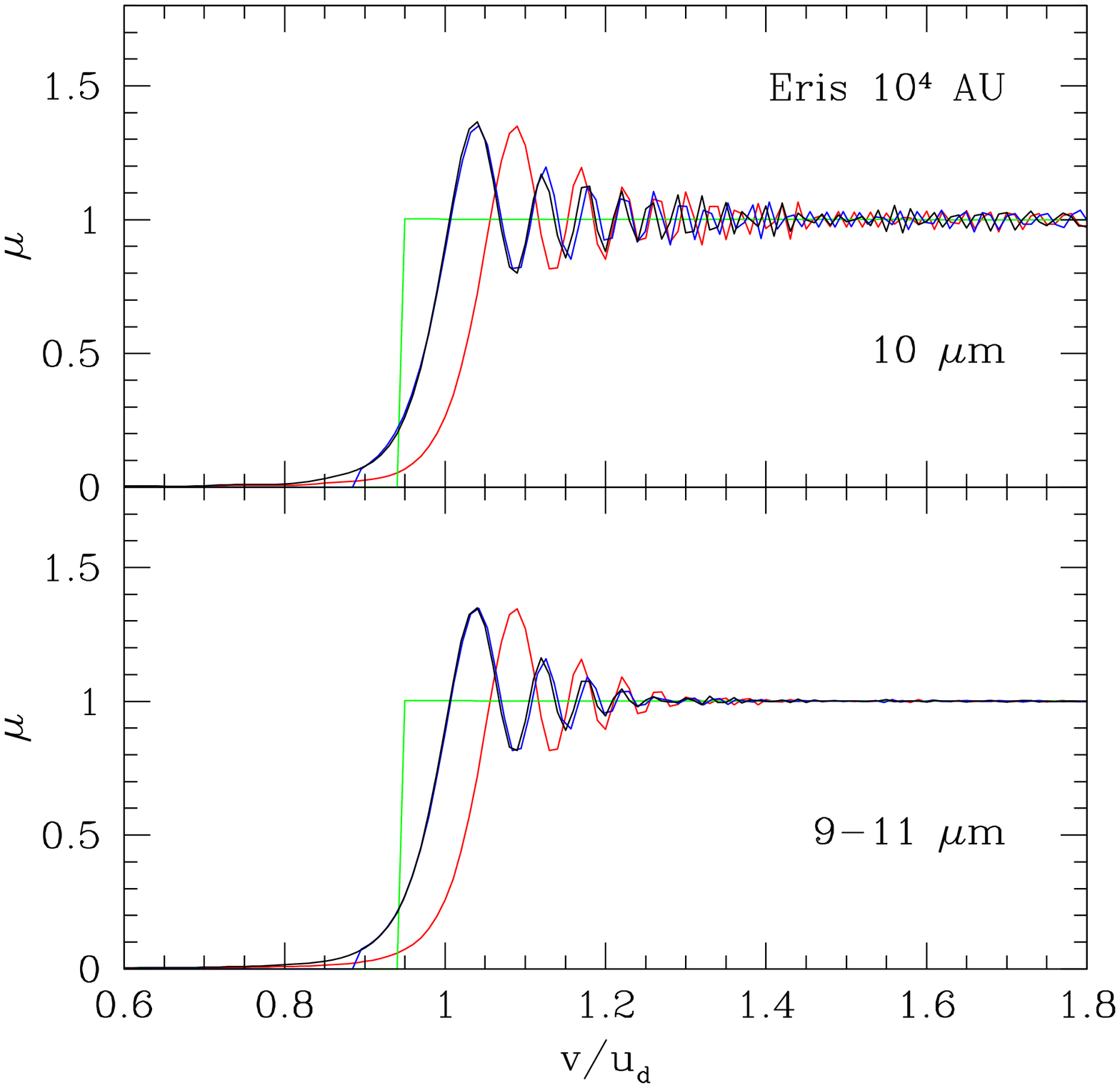}
\includegraphics[width=3.4in,clip,trim = 0 4.1in 0 0.4in]{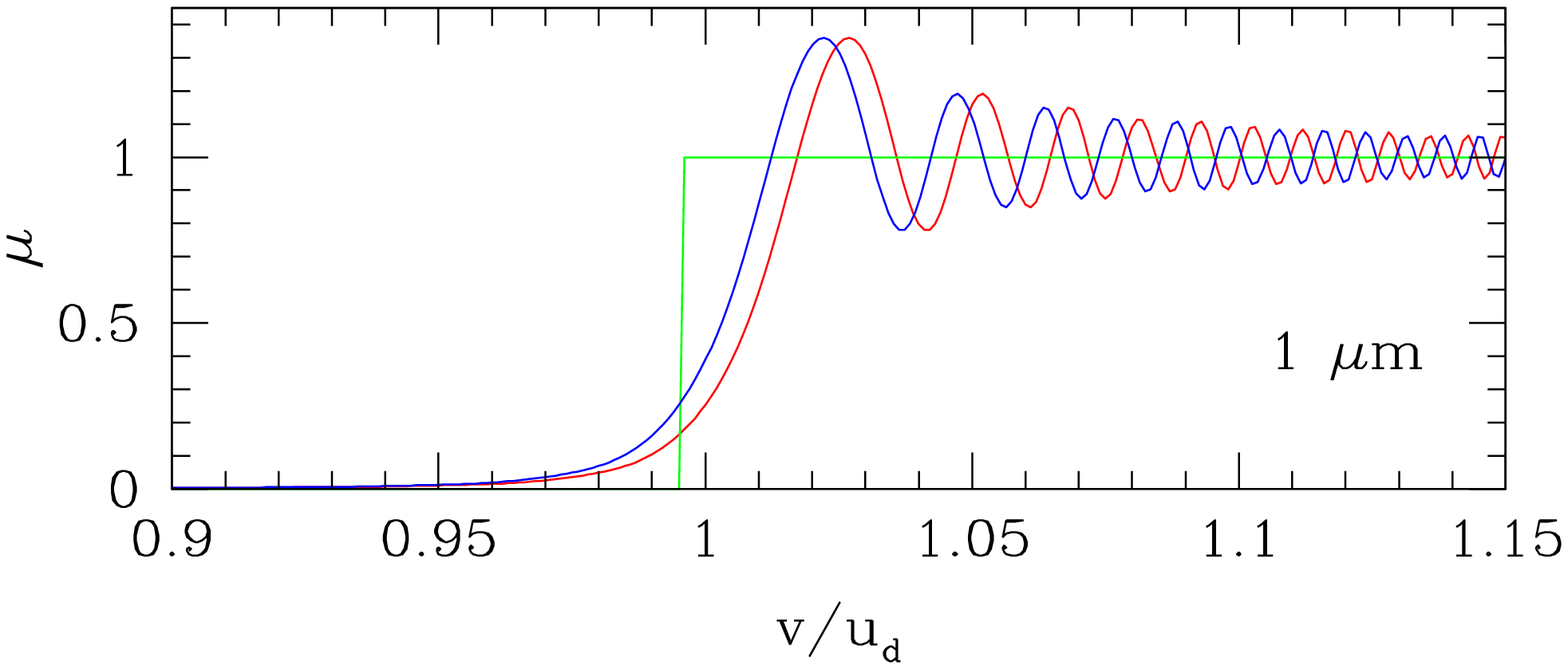}
\caption{The upper panel gives the zero bandwidth results for ten
  microns and $d=10^4$~AU (similar to the lower panel of
  Fig.~\ref{fig:1E5AU_Zero} but for $f=31.4$ and $u_d=25.$), and
  the middle panel gives the same results averaged over a
  twenty-percent bandwidth (similar to the lower panel of
  Fig~\ref{fig:1E5AU_0.2}). The $x-$axis starts at $v/u_d= 0.6$ unlike
  in the previous figures.  The blue and black curves nearly coincide.
  The lower panel gives the results for 1~$\mu$m.  At one micron the
  direct calculation including lensing becomes nearly intractable;
  however, the lensing effect is weak so the {\em a posteriori}
  lensing correction provides a good approximation.
}
\label{fig:1E4AU}
\end{figure}

Averaging the magnification over a finite bandwidth smooths out much
of the small-scale oscillation in the light curves as shown in
Fig.~\ref{fig:1E5AU_0.2}.  However, even over a moderate bandwidth of
twenty-percent the microlensed diffraction patterns are easily
distinguished from the unlensed patterns.  Again the lensing mapping
does an reasonable job of reproducing the pattern.  This important to
emphasise.  Microlensing does not simply contract the diffraction
pattern so it is not covariant with varying the relative velocities of
the lens, source and detector nor is it covariant with varying the
impact parameter of the occultation (how close to exactly aligned the
source, lens and detector become).  The lensing both increases the
amplitude of the diffractive oscillations and shifts the locations of
the peaks and troughs in a non-trivial manner, following the lensing
mapping if the lensing is weak.

If the Eris-like asteroid were closer to Earth at $10^4$~AU,
Fig~\ref{fig:regimes} indicates that the lensing would be weaker (the
lower cross is well below the red lines).  The results depicted in
Fig.~\ref{fig:1E4AU} bear this out.  The location of the first peak
moves inward from 1.09$u_d$ to 1.04$u_d$ because the light is slightly
bent about the limb of the asteroid, decreasing the duration of the
occultation by about 4.8\%, approximately the value of
$f/u_d^2\approx 5.2\%$.

\section{Conclusions}
\label{sec:conclusions}

Microlensing is important for large asteroids (similar to Eris) in the
Oort cloud and beyond; furthermore, in the near infrared and red-ward
diffraction is important to understand the light curves from the
combined microlensing and occultation of background stars by such
objects.  The effects of microlensing are not covariant with
variations in the velocity of source, lens and observer nor with
variations in the impact parameter; therefore, observations of
diffractive lensing combined with a measured distance to the asteroid
constrain the mass and radius of the asteroid, or equivalently with an
assumed value of the density of the asteroid, such observations would
yield a mass, radius and distance.  Observations of diffractive
microlensing may provide the only way of estimating the masses of such
objects unless they have a satellite as Eris does
\citep{2007Sci...316.1585B}.

The angular size of even a large asteroid such as Eris is extremely
small at the distance of the Oort cloud, so one would expect that the
chance for an occultation would be tiny.  Over the course of a year,
the asteroid will sweep out a region of the sky.  The chance of
detection each year is proportional to this area.  The semimajor axis
of the annual parallactic ellipse on the sky ranges from two to twenty
arcseconds for a distance of $10^5$ and $10^4$~AU respectively.  The
area of sky covered each year is given by the product of the arclength
around the parallactic ellipse and angular diameter of the asteroid
about 0.33~mas for an asteroid of radius 1200~km at $10^4$~AU;
therefore, on average, at the nearer distance the asteroid would cover
a region of sky 0.33 mas by 100 arcseconds or about 0.033 square
arcseconds or a fraction $6 \times 10^{-14}$ of the entire sky each
year. The solid angle covered decreases inversely with the distance
squared until microlensing starts to dominate, subsequently the area
is proportional to $d^{-3/2}$.  It is safe to assume that the asteroid
has sufficient velocity perpendicular to the Earth's orbital motion
that it moves much more that one diameter per year (much greater than
8.2 cm/s); therefore, the asteroid will cover a new region each year.
The asteroid itself may have a large proper motion that could increase
the area further --- this motion is neglected in this analysis.  Any
star within this area of sky will pass within the region $v<u_d$ in
Figs.~\ref{fig:1E5AU_Zero} through~\ref{fig:1E4AU} over the course of
that year.  If one assumes that the Oort cloud contains about ten
Earth masses of objects similar to Eris at about $10^4$~AU, one would
have to monitor about $4\times 10^8$ stars for twelve years
(twenty-four hours per day) to find a single microlensing occultation
event by an object at $10^4$~AU.  The motion of the Earth in its orbit
determines the duration of the event of about a minutes.

Currently the OGLE collaboration monitors about $4 \times 10^8$ stars
--- the present state of the art.  OGLE-III ran from 2001 to 2009 and
made nearly $2 \times 10^{11}$ photometric measurements
\citep{2008AcA....58...69U}.  The key is not only the sensitivity of
the current slew of experiments, although monitoring additional stars
or using several telescopes separated by at least the diameter of the
asteroid would increase the detection rate, but the cadence of
OGLE-III at about five minutes would simply undersample and miss such
an event.  OGLE-IV will provide both a higher sensitivity and a higher
cadence, but probably not a rapid enough cadence to distinguish such
an event.  The best approach would be a difficult hybrid of the high
cadence of observations such as \citet{2006AJ....132..819R} on a large
telescope and the monitoring of many stars as OGLE.  Because the goal
is not small bodies as in \citet{2006AJ....132..819R}, a 21ms-cadence
is overkill; a one-second cadence would be sufficient allowing
observations of much fainter stars.  Furthermore, stars with
especially small angular sizes are not necessary.  Both of these
factors along with the dedicated use of a large telescope could make
probing the largest and presumably the rarest objects in the Oort
cloud possible.

\section*{Acknowledgments}

The Natural Sciences and Engineering Research Council of Canada,
Canadian Foundation for Innovation and the British Columbia Knowledge
Development Fund supported this work.  This research has made use of
NASA's Astrophysics Data System Bibliographic Services.  I would also
like to thank the referee for many useful suggestions.

\bibliographystyle{mn2e}
\bibliography{mine,physics,math,oort}
\label{lastpage}
\end{document}